\documentclass[mathpazo]{cicp}

\usepackage{graphicx}
\usepackage{multirow}
\usepackage{caption}
\usepackage{subcaption}
\usepackage{tikz}
\usepackage{subfig}
\usepackage{subfloat}

\begin{document}
\title{Condensed Matter Laboratory: new application for quantum simulation}

\author[M.~Nakhaee]{Mohammad Nakhaee$^{1,}$\footnote{Corresponding authors: {\tt m.nakhaee@std.du.ac.ir} (M.~Nakhaee) and {\tt s.amiri@std.du.ac.ir} (S.~Amiri)} , S Ahmad Ketabi$^{1}$, Saeed Amiri$^{1,*}$ , M Ali M Keshtan$^{2}$ , Mahmoud Moallem$^{1}$ , Elham Rahmati$^{1}$, M Taher Pakbaz$^{1}$}

\address{$^{1}$Damghan University, Damghan, Iran
, $^{2}$Department of Physics, Iran University of Science and Technology, Narmak, Tehran 16844, Iran}


\begin{abstract}
The purpose of this paper is to introduce Condensed-Matter-Laboratory (CML) application for simulating solids and nanostructures and calculating different properties of them by density functional theory and using Green's function theory in tight-binding approximation to calculate phononic and spin dependent or independent electronic properties of different systems. Also, it can be used for calculating thermodynamic properties of solids and nanostructures using statistical mechanics. The CML is a cross-platform application with a graphical user interface design that is user-friendly and easy to work with. This application is written in c++ and fortran and has parallel processing ability. To show flexibility of this application some reasons are presented.
\end{abstract}

\keywords{Density functional theory, Tight-binding, Green's function, Statistical mechanics.}

\maketitle

\section{Introduction}
\label{Introduction}
Many scientists are interested in the study of nanostructures and matters in solid state, but there are many computational and mathematical challenges which cause to slow progress in this way. Due to the wideness of the problems, the existence of packages and applications will be a big help for scientific progress. Although, it can not be claimed that a package can get answer to all issues, but we can pave the way for scientists via the development of methods and algorithms. We are interested in the problems which are widely used and still really interesting. In this field, people study their favorite problems in many ways. Density functional theory is one of the most important methods in which there are several methods and algorithms, and it is still developed by theoreticians and frequently used by people. DFT is commonly applied to calculate the properties of an infinite periodic arrangement of one or more atoms (the basis) repeated at each lattice point that describes a highly ordered structure, occurring due to the intrinsic nature of its constituents to form symmetric patterns.\\
Electron density is the main player in the DFT and the main rules of the game are the two theorems of Hohenberg-Kohn and Kohn-Sham which were presented in 1964 and 1965. The accuracy of different methods returns to energy terms and different approximations which are added to the hamiltonian, like the exchange-correlation energy that contains all the interacting terms together as a single term. The rest challenge now is to find out how to express any parameters in terms of electron density and how to deal with the unknown exchange-correlation energy term compatible for any different molecules, solids, and materials.\\
One of the descriptions for XC energy in a parameterized format was done by Perdew and Zunger (1981) and Perdew and Wang (1992) known as Local Density Approximation (LDA) and for spin polarized problems Local Spin-Polarized Density Approximation (LSDA). It has been one of the most commonly used functionals in the accomplishment of the DFT before the Generalized Gradient Approximation (GGA) is formulated to generate more accurate XC functionals. Many different forms of GGA are named by their authours who have proposed theories such as PW91 (Perdew et al. 1992, Perdew and Wang 1992) and PBE (Perdew, Burke, and Ernzerhof 1996). These GGA or LDA functionals are normally well tabulated and incorporated with the pseudo-potential files.\\
Another approach which is used prevalently is tight-binding. The tight-binding or linear combination of the atomic orbitals (LCAO) method is a semi-empirical method  that is primarily used to calculate the band structure and single-particle Bloch states of a material as Slater and Koster did in 1954. The semi-empirical tight-binding method is simple and computationally very fast. It ,therefore, tends to be used in calculations of very large systems, with more than around a few thousand atoms in the unit cell. There are a number of earlier reviews ~\cite{Andersen,Ackland,Goringedag} that people working in this field should be aware of.\\
Also, we are strongly interested in calculating statistical properties of systems in both electronic and phononic cases. The connection between any quantum system and thermodynamics is possible via statistical mechanics \cite{Curado1, Curado2}. There are so many thermal properties which are interesting for experimental scientists such as mean energy, free energy, entropy and specific heat which have been formulated for bosons and fermions during many years \cite{Kardar, Isakov}.\\
The purpose of this essay is to introduce a new application for quantum simulation, it's name is Condensed Matter Laboratory (CML) and is also available for free download at {\it{http://phdstu.du.ac.ir/m-nakhaee/cml}}. Cross-platform graphical user interface of CML is written in c++ using native controls and emulates foreign functionality via wxWidgets\footnote{\it{https://www.wxwidgets.org}} tools library for GTK, MS Windows, and MacOS. Additional computational packages are written in fortran 90. Also, BLAS\footnote{\it{http://www.netlib.org/blas}} \cite{BLAS} and LAPACK\footnote{\it{http://www.netlib.org/lapack}} \cite{LAPACK} routines are used for matrix multiplications, solving systems of simultaneous linear equations and eigenvalue problems. To have high performance, portable implementation of Message-Passing-Interface (MPI) \cite{MPI} is applied to parallelize computations. For this suppose, we used MPICH\footnote{\it{https://www.mpich.org}}.\\
For the moment, this compact accumulated outline may be sufficient, and the main structure of the CML application and important topics, including the newly calculated properties, would be explained  in the rest of the paper. In each section, we report on the present status of the approaches used in detail.

\section{Construction of CML}
\label{CML}
However, variety of approaches and methods which an application uses, is a strength point and sign of ability, but having a user-friendly environment can be use to help facilitate applying those approaches and methods and handle all attributes entirely. \\

\begin{figure}[!h]
\centering
\includegraphics[scale=0.5]{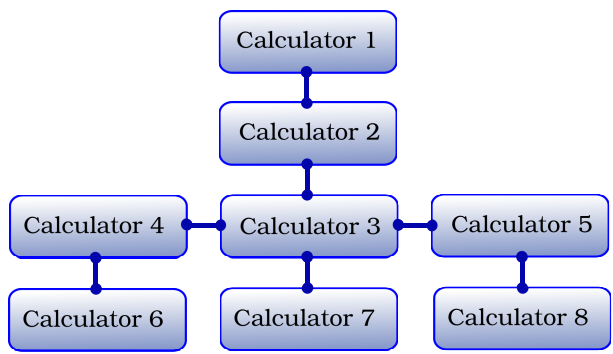}
\caption{CML project for DFT process}
\label{Projects}
\end{figure}

Assembling projects in CML is dendriform (figure \ref{Projects}) in which nodes are called calculating boxes. Calculating boxes have an input connected to parent box and some outputs connected to children. Boxes, due to their own kind, calculate some properties by inputs which comes from the parent and transmit ouputs to the children as inputs. Each of these calculating boxes makes an important contribution to our understanding of the whole project. Before considering how much CML is flexible and easy to use for the users, it is worth mentioning that, assembling a problem in such a way intercepts reduplicative calculations.

As mentioned in section \ref{Introduction}, some pre-requisites are necessary for running a typical MPI job and controlling process placement by MPICH library on a set of connected computers that work together via a machine file. It is noticeable that the {\it mpirun} command doesn't use the machine file located in the default location of the operating system. CML works inclusively to set list of hosts on which to launch MPI processes. One needs to specify IPs of the cluster's computers and also the number of nodes of each computer in settings of CML. Note that CML applys these information just in the MPICH standard. CML has different boxes with different attributes and capabilities to do quantum simulations which would be presented in the following.

\subsection{\label{Unitcell} Unit cell box}
Unit cell is used to visually simplify the crystalline patterns solids arrange themselves in. Having considered it is the least volume consuming repeating structure of any solid, it is also reasonable to look at that is the most basic calculating box for DFT calculations. One can unambiguously describes the unit cell by specifying direct lattice vectors and the set of atomic positions, and kind of atoms. The variables which are presented in table \ref{TabelUnitcell}, represent the unit cell calculating box's attributes.

\begin{table}[h]
\centering
\begin{tabular}{ |l||l| }
\hline
\multicolumn{2}{ |c| }{Unit Cell} \\
\hline \hline
\multirow{2}{*}{Unit} & unit \\
                      & lattice constant \\
\hline
\multirow{2}{*}{Unit cell group} & $ \vec{a_1} $, $ \vec{a_2} $, $ \vec{a_3} $ \\
                                 & $ \alpha $, $ \beta $, $ \gamma $ \\
\hline
\multirow{2}{*}{Unit cell atoms} & List of atoms \\
                                 & Translate to center \\
\hline
\multirow{3}{*}{Direct lattice viewport} & Driver \\
                                & Height \\
                                & Width \\
\hline
\end{tabular}
\caption{Unit Cell calculating box's attributes} \label{TabelUnitcell}
\end{table}

Viewport attribute contains some graphics drivers for different systems. One can set atoms and variables and use the unit cell options or use available groups \cite{Kovalev,Koch} (tabel \ref{TabelUnitcellGroups}) to set vectors automatically for designing a solid and view the unit cell in the viewport. Subsequently, the primitive cell will be calculated for any arbitrary unit-cell vectors.

\begin{table}[h]
\centering
\begin{tabular}{ |l|l||l|l| }
\hline
1 & Simple Cubic & 8 & Body-Centered Orthorhombic \\
\hline
2 & Face-Centered Cubic & 9 & Base-Centered Orthorhombic \\
\hline
3 & Body-Centered Cubic & 10 & Hexagonal \\
\hline
4 & Tetragonal & 11 & Rhombohedral \\
\hline
5 & Body-Centered Tetragonal & 12 & Monoclinic \\
\hline
6 & Orthorhombic & 13 & Base-Centered Monoclinic \\
\hline
7 & Face-Centered Orthorhombic & 14 & Triclinic \\
\hline
\end{tabular}
\caption{Available groups in unit cell box} \label{TabelUnitcellGroups}
\end{table}


\subsection{\label{DFT} DFT box}
Density functional theory (DFT), a theory which is currently omnipresent in our everyday computational study of atoms and molecules, solids and nano-materials, and which lies at the heart of modern many-body computational technologies. Whereas the discussion in the section \ref{Unitcell} shows that there is no use for the unit-cell calculating box on its own, the DFT box plays an important role to propel project toward other objectives are sought. DFT box gives us a bird's-eye view of the density functional theory. There is however, a further point to be considered. There is a broad range of issues lying at the foundations of DFT which would be confusing to manage all options of this powerful theory and actualy, there are several prudences and remarks that need to be noticed belonging to different solids and matters. This subject requires experience and understanding of the approaches and a much more in-depth discussion than is possible within the confines of this paper. Anyway, the DFT calculating box's attributes are presented in table \ref{TabelDFT}.

\begin{center}
\begin{table}[h]
\centering
\begin{tabular}{ |l||l| }
\hline
\multicolumn{2}{ |c| }{DFT} \\
\hline \hline
\multirow{2}{*}{Calculator} & Solution Method \\
                      & Exchange Correlation Method\\
\hline
\multirow{4}{*}{Options} & Mesh Cutoff \\
                         & Electronic Temperature \\
                         & k-grid Cutoff \\
                         & Net Charge \\
\hline
\multirow{2}{*}{Spin} & Spin Polarized \\
                      & Fix Spin \\
                      & Total Spin \\
\hline
Atomic Orbital Sets & Variables \\
\hline
Density Matrix & Options \\
\hline
\multirow{2}{*}{k-grid Monkhorst Pack} & Active \\
									 & Vectors \\
\hline
Mulliken Population Analysis & Options \\
\hline
\end{tabular}
\caption{DFT calculating box's attributes}
\label{TabelDFT}
\end{table}
\end{center}

As mentioned in section \ref{Introduction}, the second attribute (Exchange-Correlation Method) could be LDA or GGA. Regardless how LDA and GGA numerically work, LDA considers approximations for the XC energy by accounting for the varying electron densities in a system by assuming electrons see each other locally and GGA considers the electron density and its gradient at a given point. By paying more attention to what happens in actual practice before a theory is formulated, GGA would return better results than the LDA/LSDA description for XC energy. It is because of the LDA which has been becoming less popular and is often considered old-fashioned since the GGA coming, which generally describes the XC energy more accurately.\\


\subsection{\label{Bandstructure} Band Structure Box}
Band structure box refers to the electronic band structure of a solid which describes the range of energies that an electron within the solid may have and the ranges of energy that it can not have. There are different methods to calculate band structure, but in CML this box has the receptivity of the DFT, Optimization or Tight-binding boxes as a parent. Band Structure box, due to it's parent choose a compatible approach to calculate eigenenergies of the system which was defined in parrent box for each spin. Band Structure box contains properties which are presented in table \ref{TabelBandstructure}. The box needs to be initialized by the user for starting the calculation process.

\begin{table}[h]
\centering
\begin{tabular}{ |l||l| }
\hline
\multicolumn{2}{ |c| }{Band structure} \\
\hline \hline
\multirow{2}{*}{k-points} & List of k-points \\
                         & k-mesh\\
\hline
\multirow{4}{*}{Tight-binding options} & Minimum \\
                                & Maximum \\
                                & k-mesh \\
                                & Accuracy \\
\hline
\multirow{3}{*}{Reciprocal lattice viewport} & Driver \\
                                & Height \\
                                & Width \\
\hline
\multirow{3}{*}{Plotter} & Spin Up \\
                         & Spin Down \\
                         & Total \\
\hline
\end{tabular}
\caption{Band structure calculating box's attributes} \label{TabelBandstructure}
\end{table}

One should have known enough about reciprocal lattice to choose a suitable set of k-points to achieve the best results. The third attribute in table \ref{TabelBandstructure} will calculate brillouin zone for any arbitrary unit-cell vectors which are described in the parent.\\
When it is connected to DFT or Optimization boxes, this box uses siesta library to optimize the system and calculate the band structure otherwise it uses the utilities of the CML written in fortran 90 using LAPACK library which are parallelized by MPI standard. Siesta method started in 1995 \cite{Ordejon} and was coupled to several linear-scaling solvers, and coded in the Siesta program in 1996 \cite{Sanchez}, a pioneer in linear-scaling DFT.

\subsection{\label{DOS} Density of States Box}
The density of states function $ g(\epsilon) $ is defined as the number of electronic states per unit energy$ \times $volume, for electron energies at energy $ \epsilon$. Comparatively, in consonance with electrons $ g(\omega) $ is defined as the number of phononic modes, for phonon modes at frequency $ \omega$. The density of states function is important for calculations of the effects based on the band theory like transition probability \cite{Harrison}. The actual transition probability depends on how many states are available in both the initial and final energies. The band structure is not a reliable guide here, since it only tells about the bands along high symmetry directions. What one needs is the full density of states across the whole Brillouin zone, not just the special directions and has to sample the Brillouin zone evenly. Also, it appears in calculations of conductivity where it presents the number of mobile states, and in computing scattering rates where it presents the number of final states after scattering in both of electrical and phononic cases \cite{kuno,ashcroft}. \\
The parent of this calculating box could be Optimization, Tight-binding or Band structure boxes. With this comment, it is noteworthy to mention that the approach to calculate the density of states depends on the parent's kind chosen by the user. DOS box contains properties presented in table \ref{TableDOS}.

\begin{table}[h]
\centering
\begin{tabular}{ |l||l| }
\hline
\multicolumn{2}{ |c| }{Density of states} \\
\hline \hline
\multirow{4}{*}{Options} & Minimum Energy \\
                         & Maximum Energy\\
                         & Number of points\\
                         & Broadening\\
\hline
\multirow{3}{*}{Plotter} & Spin Up \\
                         & Spin Down \\
                         & Total \\
\hline
\end{tabular}
\caption{DOS calculating box's attributes} \label{TableDOS}
\end{table}

DOS as a function of energy can be calculated by doing a sum of all bands over all energies of k-vectors, using the following equation:

\begin{equation} \label{eqdos}
      g(\epsilon) = \frac{1}{N_k}\sum_k\delta(\epsilon-\epsilon_k)
\end{equation}

where $ N_k $ is the total number of k-vectors covering the Brillouin zone. Simply using the delta
function as a counter gives a histogram for the DOS.\\
In the tight-binding approximation, this box needs the hamiltonian of the system (the Tight-Binding calculating box outputs) to calculate the density of states $ g(\epsilon) $ of the system as follows:

\begin{equation} \label{eqdos}
      g(\epsilon) = -\frac{1}{\pi}Im Tr(\hat{G})
\end{equation}


\subsection{\label{Tightbinding} Tight-binding Box}
Semi-empirical tight-binding method is an acceptable approximation to hide daintily all information about eigenvectors behind a concept which is called hopping. The tight-binding description eventually has progressed to treat systems containing up to thousands of atoms and made computer experiments possible for many structures. Proponents of tight-binding, have also developed several computational and theoretical methods to make computations faster. These days, the methods have been so successful in both accuracy and efficiency that scientists are using them routinely in a wide range. In CML, one is able to assemble three kinds of systems: Junctions, ribbons and molecules to study. A junction system can be defined by the following hamiltonian:
\begin{align}\label{SystemLinear}
H =
\left(
\begin{array}{ccccccc}
\ddots    &  \ddots    &  \bold{0}  &  \bold{0}  &  \bold{0}  &  \bold{0}  &  \bold{0}  \\
\ddots    &  H_{L11}   &  V_{L10}   &  \bold{0}  &  \bold{0}  &  \bold{0}  &  \bold{0}  \\
\bold{0}  &  V_{L10}^\dagger   &  H_{L00}   &  V_{LC}    &  \bold{0}  &  \bold{0}  &  \bold{0}  \\
\bold{0}  &  \bold{0}  &  V_{LC}^\dagger    &  H_{C}     &  V_{CR}    &  \bold{0}  &  \bold{0}  \\
\bold{0}  &  \bold{0}  &  \bold{0}  &  V_{CR}^\dagger    &  H_{R00}   &  V_{R01}   &  \bold{0}  \\
\bold{0}  &  \bold{0}  &  \bold{0}  &  \bold{0}  &  V_{R01}^\dagger   &  H_{R11}   &  \ddots    \\
\bold{0}  &  \bold{0}  &  \bold{0}  &  \bold{0}  &  \bold{0}  &  \ddots    &  \ddots
\end{array}
\right)
\end{align}

in which $ H_\alpha $ is the hamiltonian of the part $ \alpha $ (Center, Left or Right) and $ V_{\alpha \beta} $ is the coupling matrix between parts $ \alpha $ and $ \beta $.
Also, for a ribbon we can define the hamiltonian as fallows:
\begin{align}\label{SystemLinear}
H =
\left(
\begin{array}{ccccc}
\ddots    &  \ddots    &  \bold{0}  &  \bold{0}  &  \bold{0}  \\
\ddots    &  H_{C}     &  V         &  \bold{0}  &  \bold{0}  \\
\bold{0}  &  V^\dagger &  H_{C}     &  V         &  \bold{0}  \\
\bold{0}  &  \bold{0}  &  V^\dagger &  H_{C}     &  \ddots    \\
\bold{0}  &  \bold{0}  &  \bold{0}  &  \ddots    &  \ddots
\end{array}
\right)
\end{align}

for a molecule there is no coupling and the system can be defined by $H = H_{C}$. In the case of spin dependent electron, we need a standard for CML application in which we suppose that the hamiltonian should be written as follows:

\begin{align}\label{SystemLinear}
H =
\left(
\begin{array}{cc}
  H_{\alpha}^{\uparrow}  &  V_{\alpha\alpha}^{\uparrow\downarrow}  \\
  V_{\alpha\alpha}^{\downarrow\uparrow}  &  H_{\alpha}^{\downarrow}  \\
\end{array}
\right)
\end{align}

\begin{table}[h]
\centering
\begin{tabular}{ |l||l| }
\hline
\multicolumn{2}{ |c| }{Tight-binding} \\
\hline \hline
\multirow{2}{*}{Calculator} & Electronic\\
                            & Phononic \\
\hline
\multirow{2}{*}{Options} & Spin Polarized \\
                         & Kind of System\\
\hline
\multirow{7}{*}{System} & Left lead Matrix\\
                         & Center Matrix\\
                         & Right Lead Matrix\\
                         & Left-Center Coupling Matrix\\
                         & Center-Right Coupling Matrix\\
                         & Inner Left Coupling Matrix\\
                         & Inner Right Coupling Matrix\\
\hline
\end{tabular}
\caption{Tight-binding calculating box's attributes} \label{TableTightbinding}
\end{table}

All we need to define a tight-binding system comes in Table \ref{TableTightbinding}. One needs to set values and put matrices in the condition appropriate to the start of defining system.

\subsection{\label{dispersion} Phonon Dispersion Box}
The study of quantum of vibrational energy (phonon) is an important part of condensed matter physics. The phonon has energy $ \hbar \omega $ and wave vector $ q $ with the frequency $ \omega_\mu(q)$ as a function of $ q $ known as the phonon dispersion relation in which $ \mu $ refers to the different branches corresponding to the different frequency modes. The phonon dispersion relation is periodic in the reciprocal lattice, as that of the electron, when a reciprocal lattice vector is added to the $ q $. All unique solutions are then obtained if $ q $ is restricted to a cell in the reciprocal space which is usually chosen to be the first Brillouin zone as the primitive cell. Phonons play a major role in many of the physical properties of condensed matter, like thermal conductivity and electrical conductivity. To calculate the dispersion of a periodic system we need to calculate eigenvalues of the equation

\begin{equation} \label{MatTtExpInt}
      D(q)\psi(q) = \omega_q^2\psi(q)
\end{equation}

In which $D(q)$ is the $ q $ dependent dynamical matrix of the system \cite{ashcroft}.
\begin{equation} \label{MatTtExpInt}
      D(q) = D_c + V_{CR} e^{-i q R} + V_{LC} e^{i q R}
\end{equation}

In which $D_c$ is the dynamical matrix of the center part and $V_{CR}$ is the coupling matrix between the center and right unit cell. It should be noted that the hamiltonian must be an hermitian matrix and so $V_{CR}$ is equal to $V_{LC}^\dagger$. The Phonon-dispersion box contains attributes presented in table \ref{TablePhononDispersion}.

\begin{table}[h]
\centering
\begin{tabular}{ |l||l| }
\hline
\multicolumn{2}{ |c| }{Phonon Dispersion} \\
\hline \hline
\multirow{3}{*}{Options} & Minimum \\
                         & Maximum \\
                         & Number of points\\
\hline
\end{tabular}
\caption{Phonon Dispersion calculating box's attributes} \label{TablePhononDispersion}
\end{table}

By assembling the dynamical matrix in Tight-binding box by setting the attribute Calculator \ref{TableTightbinding} on phononic case, one is able to connect this box as a child of that. It is notable that, the system of parent \ref{TableTightbinding} must be set on the Ribbon.

\subsection{\label{Current} Current Box}
Understanding electricity is needed to know a wide variety of well-known effects, such as lightning, static electricity and electromagnetic induction. One of them is an electric current which is a flow of electric charge. Transmission through a junction has been interesting for scientists to observe how electrons produce electric current when a bias voltage is applied. Electronic or phononic current flowing from left to right through a two-lead junction at different equilibrium potentials is given by the Landauer formula as follows \cite{Datta}:

\begin{equation} \label{eqcurr}
      I =\int \frac{\hbar \omega}{2 \pi} T(\omega)(f_L - f_R)d\omega
\end{equation}

This box's codes are written in fortran 90 using LAPACK library and it is able to calculate in parallel mode. Current Box contains some properties which are presented in table \ref{TableCurrent}.

\begin{table}[h]
\centering
\begin{tabular}{ |l||l| }
\hline
\multicolumn{2}{ |c| }{Current} \\
\hline \hline
\multirow{3}{*}{Options} & Minimum Voltage \\
                         & Maximum Voltage \\
                         & Number of points\\
\hline
\multirow{5}{*}{Transmission} & Minimum Energy \\
                         & Maximum Energy \\
                         & Number of points\\
                         & Broadening\\
                         & Accuracy\\
\hline
\multirow{3}{*}{Plotter} & Spin Up \\
                         & Spin Down \\
                         & Total \\
\hline
\end{tabular}
\caption{Current calculating box's attributes} \label{TableCurrent}
\end{table}


\subsection{\label{StatisticalMechanics} Statistical Mechanics Box}
In the grand canonical ensemble, the density operator $ \hat{\rho} $ operates on a Hilbert space with an indefinite number of particles. The density operator must ,therefore, commute not only with the Hamiltonian operator $ \hat{H} $ but also with a number operator $ \hat{n} $ whose eigenvalues are different for fermions and bosons. The precise form of the operator $ \hat{\rho} $ can be obtained as follows:

\begin{equation} \label{eqSM1}
	\hat{\rho} = \frac{1}{\mathcal{Z}} e^{-\beta (\hat{H} - \mu \hat{n})}
\end{equation}

in which the grand canonical partition function $ \mathcal{Z} $ applies to a grand canonical ensemble and it would be defined in this way:

\begin{equation} \label{eqPartitionFunction}
	\mathcal{Z} = Tr(e^{-\beta (\hat{H} - \mu \hat{n})})
\end{equation}

$ \mathcal{Z} $ will be expanded in fermionic case and bosonic case differently. For systems of identical femions and identical bosons, an exchange of particles does not change the physical state. Therefore, the factor $ g({f_{n m}}) $ is just 1 for both kinds of systems. Moreover, the occupation number of a state characterized by $n$ for a bosonic system would be any number between $0$ and $N$:

\begin{equation} \label{eqSM1}
      g_{n_\epsilon} = 0,1,2,...,N
\end{equation}

For fermions, the Pauli exclusion principle forbids two identical particles from occupying the same quantum state. This restricts the occupation numbers to be either 0 or 1: 

\begin{equation} \label{eqSM2}
      g_{n_\epsilon} = 0,1
\end{equation}

In consonance with definitions of the average energy $ \mathcal{U} $, entropy $ \mathcal{S} $, free energy $ \mathcal{F} $ and specific heat $ C_v $, the most important thermodynamic and statistical properties of the system can be written compactly as equations \ref{eqSMFormula} \cite{Kardar,Pathria}.

\begin{align}  \label{eqSMFormula}
      &\mathcal{U} = Tr(\hat{\rho}.\hat{H}) \nonumber \\
      &\mathcal{S} = k T (\frac{\partial ln \mathcal{Z}}{\partial T}) \nonumber \\
      &\mathcal{F} = \mathcal{U} - \mathcal{S} T \nonumber \\
      &C_v = (\frac{\partial \mathcal{U}}{\partial T})_v
\end{align}

\begin{table}[h]
\centering
\begin{tabular}{ |l||l| }
\hline
\multicolumn{2}{ |c| }{Statistical Mechanics} \\
\hline \hline
\multirow{4}{*}{Options} & Minimum Temperature \\
                         & Maximum Temperature \\
                         & Number of points\\
                         & Accuracy\\
\hline
\multirow{7}{*}{Plotter} & Energy \\
                         & Free Energy \\
                         & Entropy \\
                         & Specific Heat \\
                         & Spin Up \\
                         & Spin Down \\
                         & Total \\
\hline
\end{tabular}
\caption{Statistical Mechanics calculating box's attributes} \label{TableStatisticalMechanics}
\end{table}

Statistical Mechanics box contains attributes presented in table \ref{TableStatisticalMechanics} in which spin-dependent attributes would belong to fermionic systems. This box can choose Band structure, Phonon dispersion and DOS calculating boxes as it's parent to get inputs. The Band structure and Phonon dispersion boxes give it eigenvalues of the system as inputs to calculate partition function by the equation \ref{eqPartitionFunction} via fermionic or bosonic approach (due to the parent kind). If one connects this box to the DOS calculating box, the method to calculate partition function would be different. It should be noted that, the canonical partition function $ Q(\beta) $ is just the Laplace transform of the density of states $ g(\epsilon) $ (see equation \ref{eqLaplaseTransform}) \cite{Pathria}. This is a good flight to the statistical world via the density of states of any system.

\begin{equation} \label{eqLaplaseTransform}
      Q(\beta) =\int g(\epsilon)e^{-\beta \epsilon}d\epsilon
\end{equation}

\subsection{\label{Optimization} Optimization Box}
The problem is that, one may need to find the atomic positions which would have to be before starting DFT process. Before proceeding further, it is noticeable that applying a proper computational method is very critical to produce a better atomic structure. The method using this classical mechanics for atomic movements is called molecular dynamics (MD). It initially simulates a handful of molecules. By using atoms as the lowest level of information, MD will be treated for the predictions of static and dynamic properties of materials at the introductory level. Attributes of the Optimization box are presented in table \ref{TableOptimization}.

\begin{table}[h]
\centering
\begin{tabular}{ |l||l| }
\hline
\multicolumn{2}{ |c| }{Optimization} \\
\hline \hline
Optimizer & Method \\
\hline
\multirow{7}{*}{Options} & Variable Cell \\
                         & Constant Volume\\
                         & Relax Cell Only\\
                         & Broyden Cycle on Max it\\
                         & Quench \\
                         & Fire Quench \\
                         & Aneal Option\\
\hline
\multirow{7}{*}{Inputs} & Itteration \\
                         & Max Force Tolerance \\
                         & Max Stress Tolerance \\
                         & Max CG Displ \\
                         & Precondition Variable Cell \\
                         & History Steps\\
                         & Variables\\
\hline
 Outputs & Options \\
\hline
\multirow{5}{*}{Plotter} & Total Energy \\
                         & KS Energy \\
                         & Volume \\
                         & Pressure \\
                         & Temperature \\
\hline
\end{tabular}
\caption{Optimization calculating box's attributes} \label{TableOptimization}
\end{table}

However, there are a lot of applications and methods to find the optimum energy of the molecular dynamic of a system (ie. Gaussian program\footnote{\it{http://www.gaussian.com}}, MondoSCF package\footnote{\it{http://www.t12.lanl.gov/home/mchalla/MondoSCF.html}}, QChem program \cite{Shao} and etc.), we used siesta package to relax the unit cell.

\subsection{\label{Transmission} Transmission Box}
Sometimes we need to study the properties of a system by the Green's function method. In this method, the most important property of physical observables in transport, such as thermal or energy current, is the transmission. The transport of electrons or phonons in a medium having negligible electrical resistivity is caused by scattering where the system is assumed to be ballistic. Ballistic conduction is not limited to electrons but can also apply to phonons. The tight-binding description of transmission through a junction has been interesting for scientists to observe how much a wave packet would transmit in a specific energy. This box would calculate both electronic and phononic transmission and the density of states. In this way, we can also calculate the density of states $ g(\epsilon) $ of each part of system by equation \ref{eqdos}. This box needs the hamiltonians of all parts of the system which the Tight-binding calculating box gives it them as it's parent. All attributes one needs to control this box are presented in table \ref{TableTransmission} and also this box is able to calculate in parallel mode.

\begin{table}[h]
\centering
\begin{tabular}{ |l||l| }
\hline
\multicolumn{2}{ |c| }{Transmission} \\
\hline \hline
\multirow{6}{*}{Options} & Bias Voltage \\
                         & Minimum Energy \\
                         & Maximum Energy\\
                         & Number of points\\
                         & Broadening\\
                         & Accuracy\\
\hline
\multirow{7}{*}{Plotter} & Transmission \\
                         & Center DOS \\
                         & Left DOS \\
                         & Right DOS \\
                         & Spin Up \\
                         & Spin Down \\
                         & Total \\
\hline
\end{tabular}
\caption{Transmission calculating box's attributes} \label{TableTransmission}
\end{table}

The concept, which this box would use to calculate transmission $ T(\epsilon) $ is the Landauer-B{\"u}ttiker formalism via Green's function approach as follows:

\begin{equation} \label{eqSM2}
      T(\epsilon) = Tr(\hat{\Gamma}_L.\hat{G}.\hat{\Gamma}_R.\hat{G}^\dagger)
\end{equation}

in which $ G $ is the retarded green function of the system and $\Gamma_\alpha$ is the so-called broadening function of lead $ \alpha $ ($ \alpha $ = Left or Right). It is given in terms of self-energy of the lead $ \Sigma_\alpha $ as equation \ref{eqGamma} \cite{Datta}.

\begin{equation} \label{eqGamma}
      \hat{\Gamma}_\alpha = i(\hat{\Sigma}_\alpha - \hat{\Sigma}_\alpha^\dagger)
\end{equation}

In the case of phonon transmission, if the center is ballistic, the formula can be further simplified. The result for the ballistic system is called Landauer formula with a transmission function as equation \ref{eqCaroli} known as the Caroli formula.

\begin{equation} \label{eqCaroli}
      T(\omega) = Tr(\hat{G}^r.\hat{\Gamma}_L.\hat{G}^a.\hat{\Gamma}_R)
\end{equation}

This formula is an efficient way of computing the quantum thermal transport in terms of the Green's functions used on treating transport by NEGF which is that of Caroli, et al. \cite{Wang2006,Wang2008}.

\subsection{\label{SuperCell} Super Cell Box}
When one needs to repeat the unit-cell for a finite number at the desire direction this calculating box has some attributes (Table \ref{TableSuberCell}) to do that. This calculating box contains three vectors and could be connected to the unit-cell box as a child. After setting this box the project cloud be propelled toward other objectives sought via a DFT box.

\begin{table}[h]
\centering
\begin{tabular}{ |l||l| }
\hline
\multicolumn{2}{ |c| }{Super Cell} \\
\hline \hline
Vectors & $ \vec{V_1} $, $ \vec{V_2} $, $ \vec{V_3} $ \\
\hline
\end{tabular}
\caption{Super Cell calculating box's attributes} \label{TableSuberCell}
\end{table}

\section{Examples}
\label{Examples}
\begin{figure}[!htbp]
\centering
\includegraphics[scale=0.5]{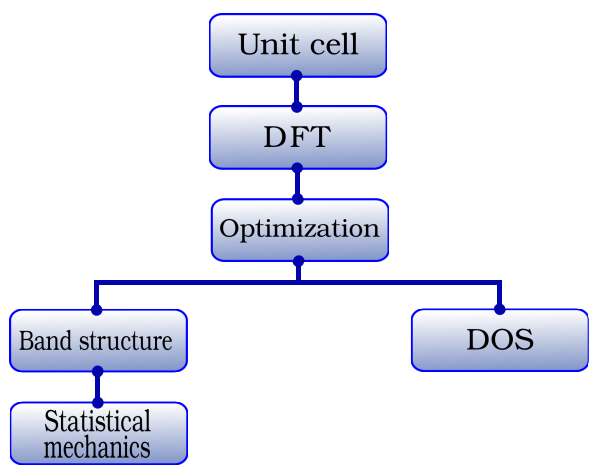}
\caption{CML project for DFT process}
\label{PhosphoreneProjects}
\end{figure}

There is an old Latin saying: {\it{Longum iter est per praecepta, breve et efficax per exempla.}} (It's a long way by the rules, but short and efficient with examples). In this section, we will follow some applied and typical examples and see how everything we learned in the previous section works. To show flexibility of the CML application and how much it is simple, we will apply two methods: DFT and tight-binding for three examples: electrical properties of Phosphorene surface by DFT and electrical properties of a Graphene nanoribbon in tight-binding approximation and phononic properties of a typical chain (contains two different atoms as atomic basis). The project for Phosphorene could be easily assembled just like the figure \ref{PhosphoreneProjects}. The unit-cell box contains atomic positions of the Phosphorene surface. By connecting it to the DFT box, the density functional theory will set up by default values. If we have the exact locations there is no need to setup the Optimization box, but here we assume it is needed. To compute the band structure and the density of states one could just add their boxes as shown in figure \ref{PhosphoreneProjects}.\\

\begin{figure}[!htbp]
\centering
	\begin{subfigure}[b]{.3\textwidth}
  		\centering
		\includegraphics[scale=0.37]{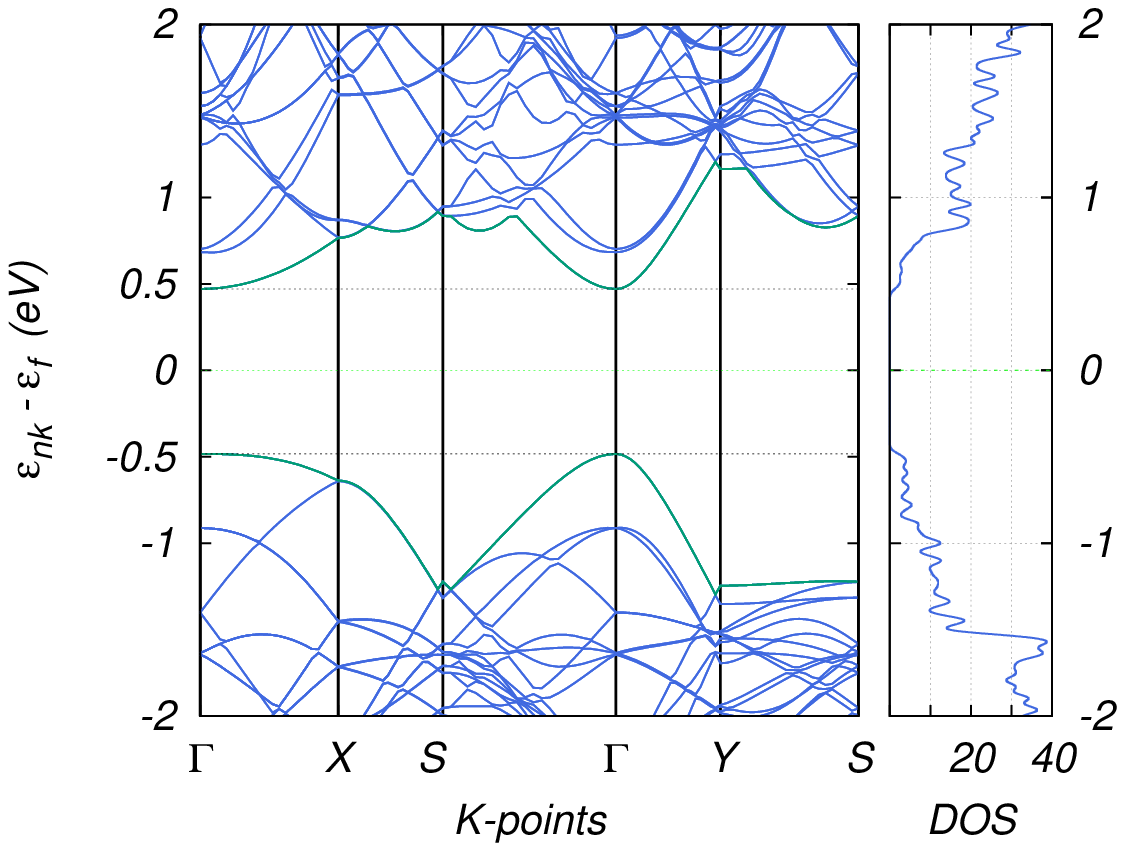}
		\caption{}
		\label{PhosphoreneBS}
	\end{subfigure}
	\begin{subfigure}[b]{.3\textwidth}
  		\centering
		\includegraphics[scale=0.37]{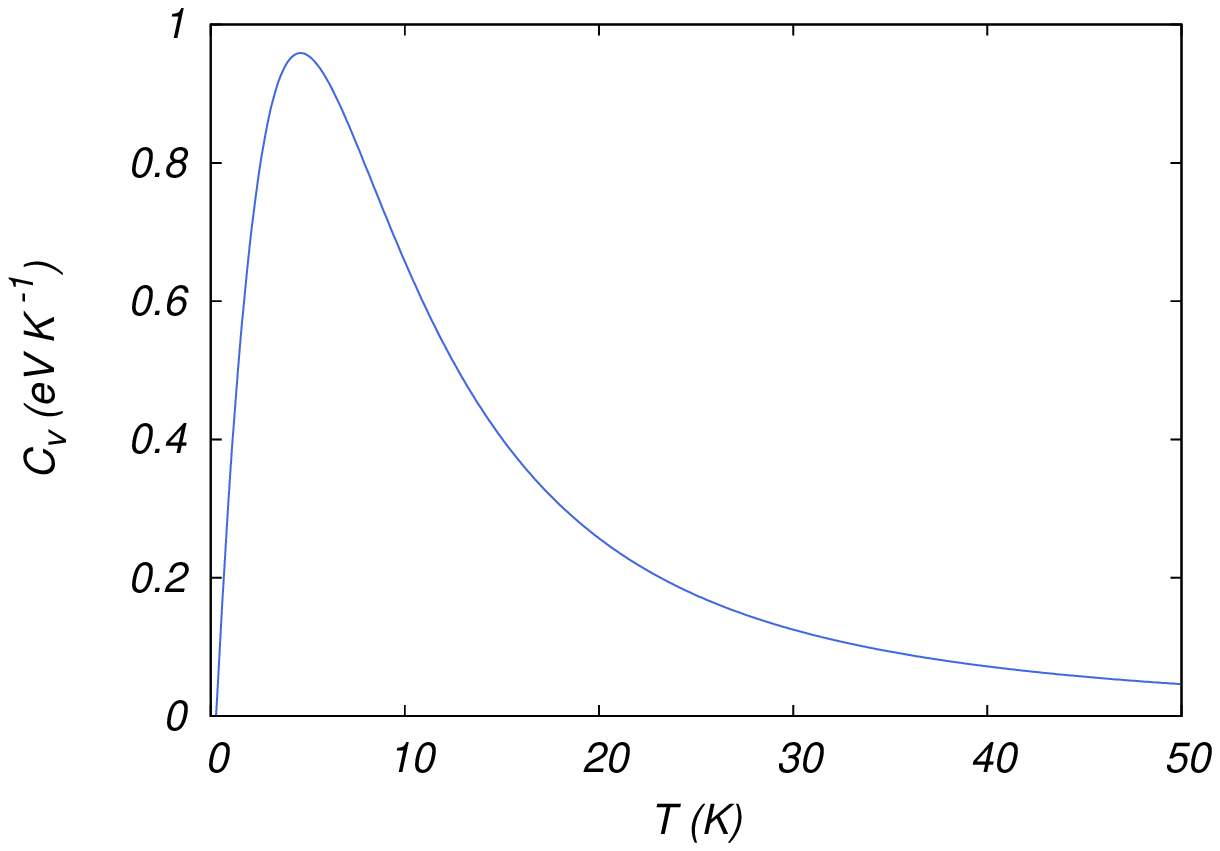}
		\caption{}
		\label{PhosphoreneCv}
	\end{subfigure}		
	\begin{subfigure}[b]{.3\textwidth}
  		\centering
		\includegraphics[scale=0.37]{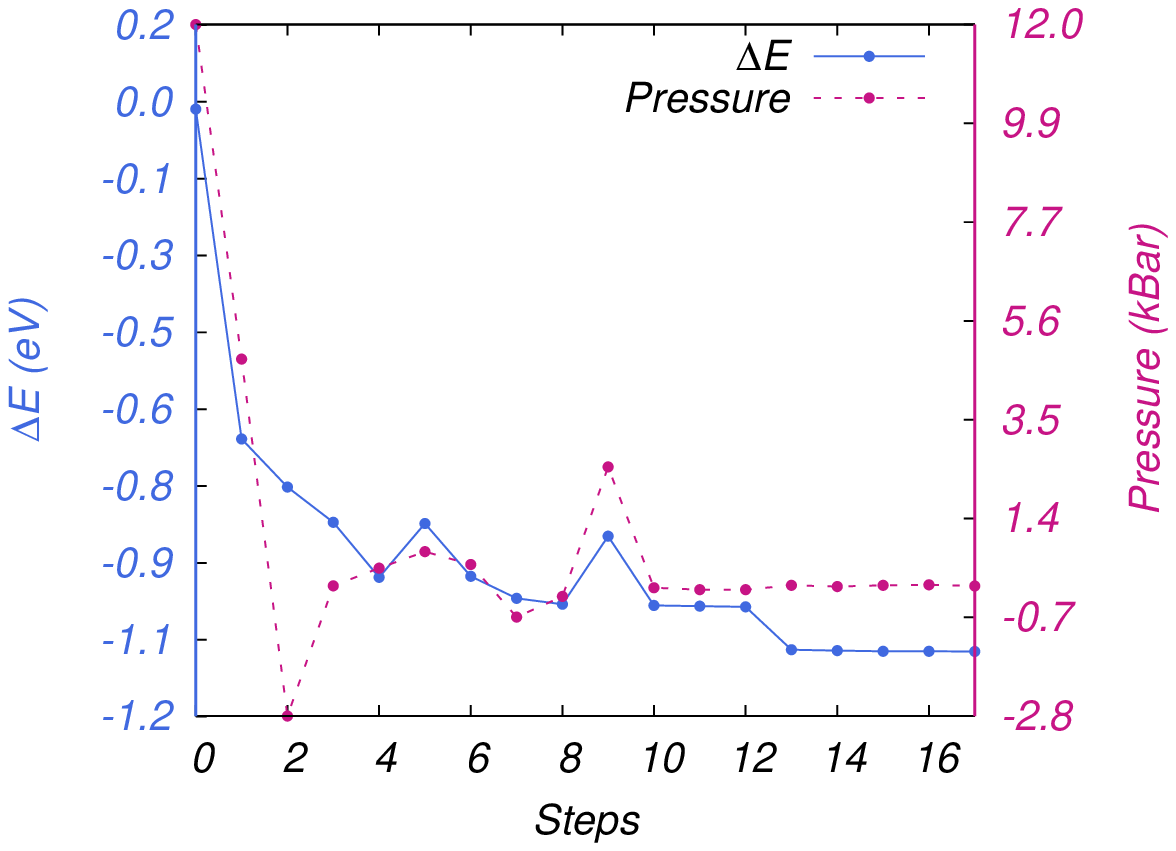}
		\caption{}
		\label{PhosphoreneOpt}

	\end{subfigure}
\caption{(Colour online) Electronic properties of Phosphorene surface: (a)Energy bands and DOS of Phosphorene, (b)Electronic specific heat of Phosphorene, (c)Molecular dynamic's total energy and pressure.}
\label{PhosphoreneProperties}
\end{figure}

After starting the process one can follow the progress of molecular dynamics by the plotter attribute of the Optimization box. It is strongly recommended to check energies if the system can converge or no. Figure \ref{PhosphoreneOpt} shows the variety of the total energy and the pressure by steps. Also, the band-structure box contains the k-paths which are set by the user. After completing the run, the density of states and the band structure of the Phosphorene can be plotted respectively by plotter attributes of the DOS and Band-structure calculating boxes as shown in figure \ref{PhosphoreneBS}. Also, one may need to add the Statistical-Mechanics box for finding out the electronic specific heat of Phosphorene. (figure \ref{PhosphoreneCv})\\

\begin{figure}[!htbp]
\centering
\includegraphics[scale=0.5]{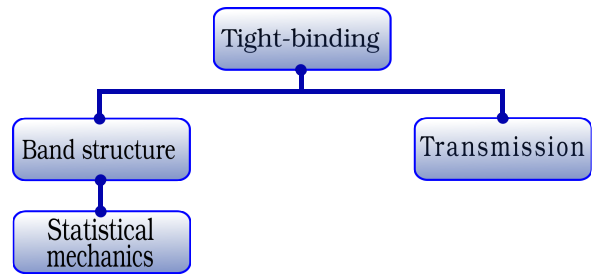}
\caption{CML project to calculate electronic properties of a Graphene nanoribbon}
\label{NanoRibbonProjects}
\end{figure}

Figure \ref{NanoRibbonProjects} shows a CML project to calculate electronic properties of a Graphene nanoribbon. The first calculating box contains hamiltonians of the system and other boxes can easily connect to their parrents to calculate transmission, current and statistical properties of the nanoribbon.

\begin{figure}[!htbp]
\centering
	\begin{subfigure}[b]{.3\textwidth}
  		\centering
		\includegraphics[scale=0.37]{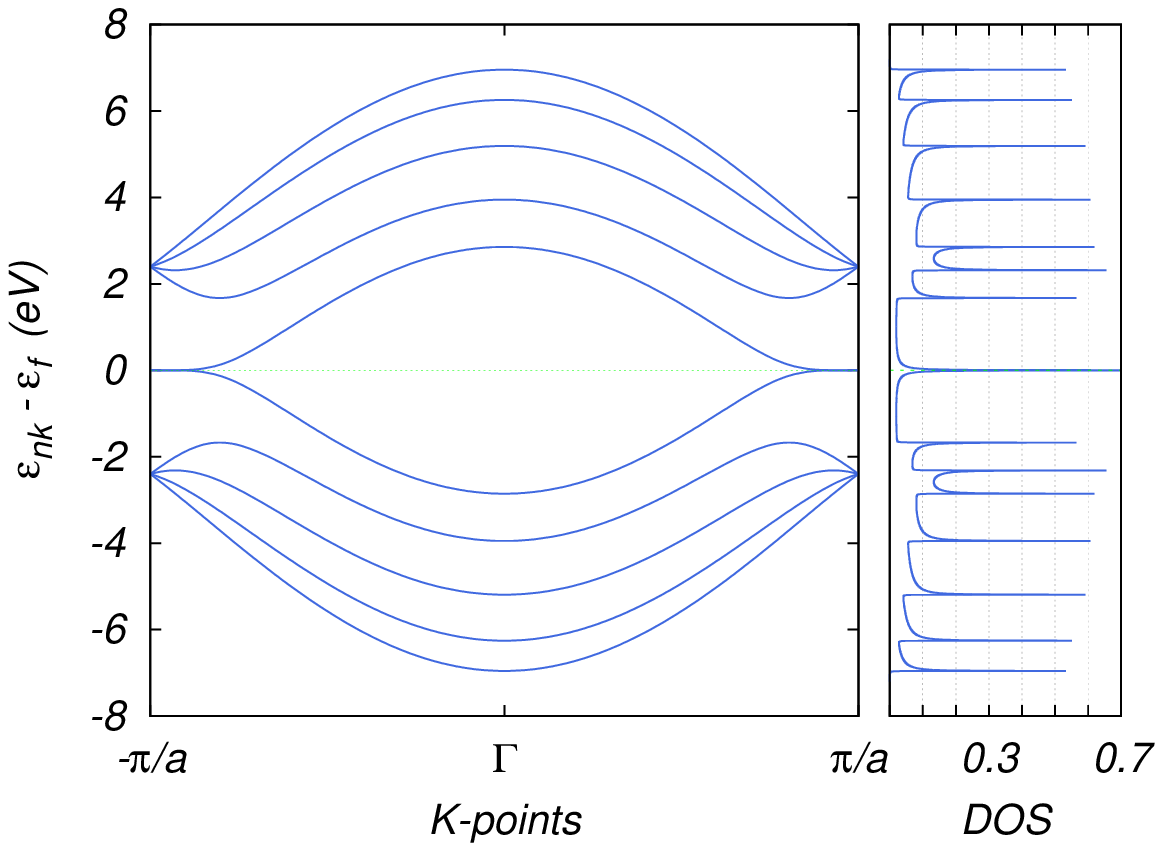}
		\caption{}
		\label{GrpheneBands}
	\end{subfigure}
	\begin{subfigure}[b]{.3\textwidth}
  		\centering
		\includegraphics[scale=0.37]{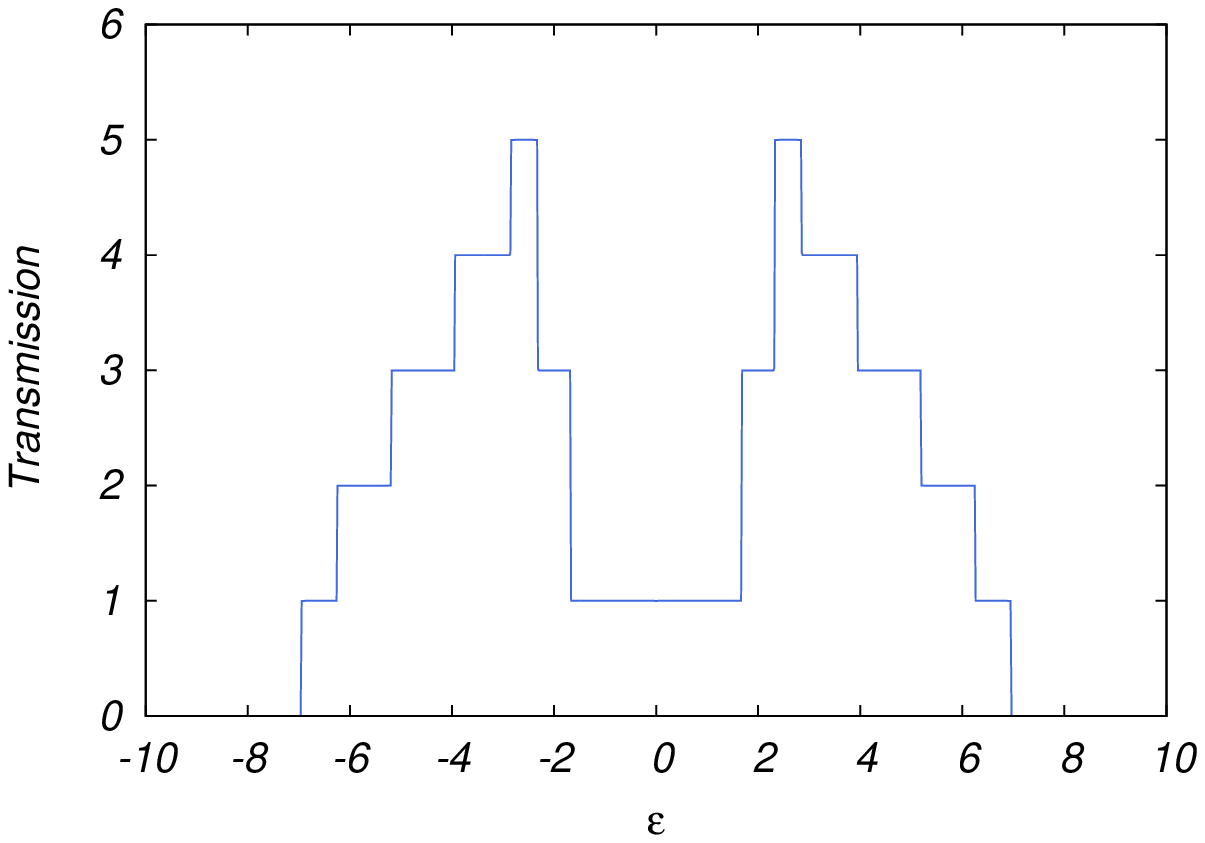}
		\caption{}
		\label{GrapheneTransmission}
	\end{subfigure}		
	\begin{subfigure}[b]{.3\textwidth}
  		\centering
		\includegraphics[scale=0.37]{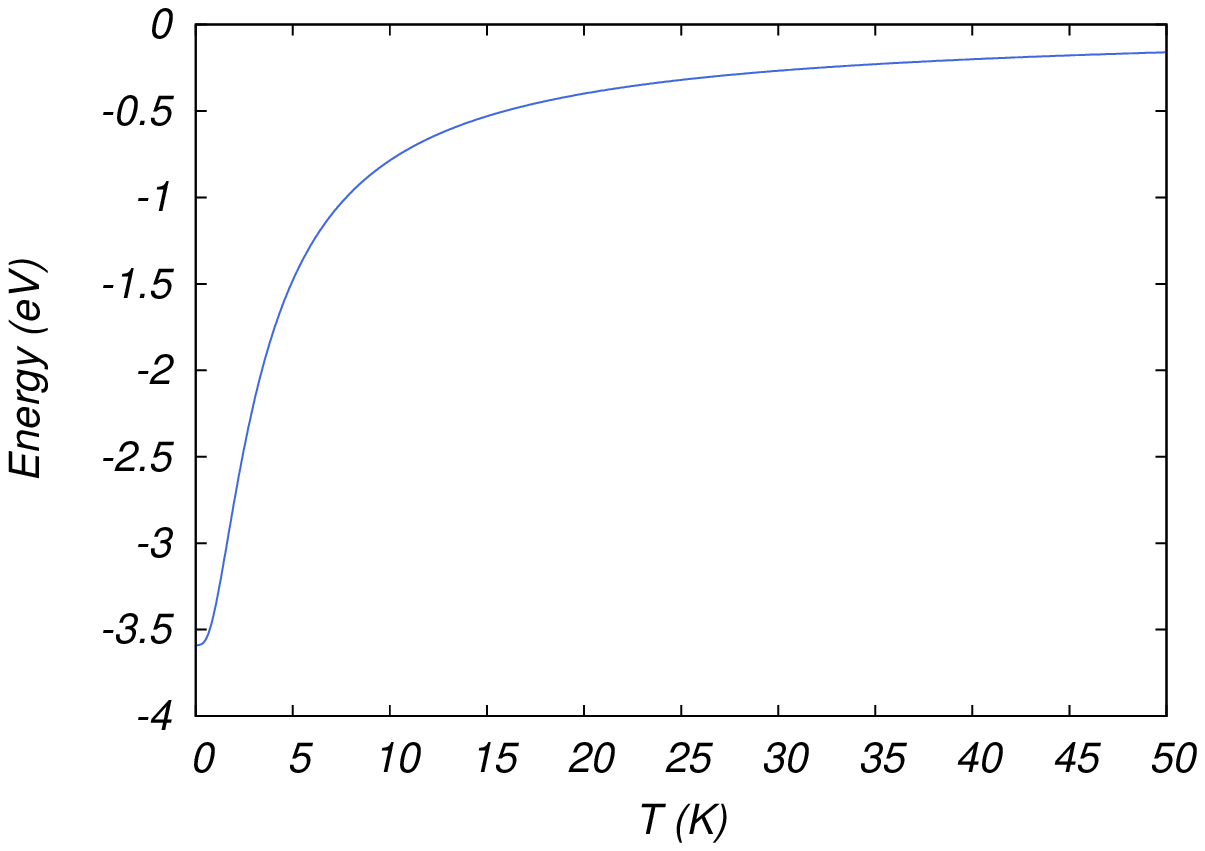}
		\caption{}
		\label{GrapheneEnergy}

	\end{subfigure}
\caption{(Colour online) Electronic properties of Graphene nanoribbon: (a)Energy bands and DOS of Graphene nanoribbon, (b)Transmission of Graphene nanoribbon, (c)Mean energy of Graphene nanoribbon.}

\label{PhosphoreneProperties}
\end{figure}

Figure \ref{GrapheneTransmission} shows the transmission of the nanoribbon. Note that when we talk about nanoribbon as a junction, it means that the leads are made of the same matterial which are used in center part. So the hamiltonians of the left and the right leads must be exactly the same as the center one. Also, it should be noted that, the system attribute shown in Table \ref{TableTightbinding} is different for calculating band structure and others. For band structure calculating, it must be set on nanoribbon and for others, it has to be a junction. Figure \ref{GrpheneBands} shows energy bands of graphene nanoribbon and the density of states side by side. Also, statistical mean energy of electrons in a graphene nanoribbon is shown in figure \ref{GrapheneEnergy}.

\begin{figure}[!htbp]
\centering
\includegraphics[scale=0.5]{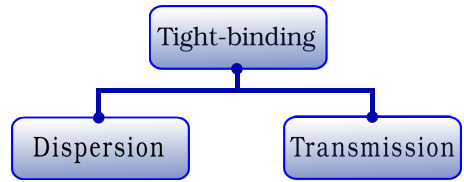}
\caption{CML project to calculate phononic dispersion of a chain.}
\label{PhononChainProjects}
\end{figure}

\begin{figure}[!htbp]
\centering
	\begin{subfigure}[b]{.3\textwidth}
  		\centering
		\includegraphics[scale=0.37]{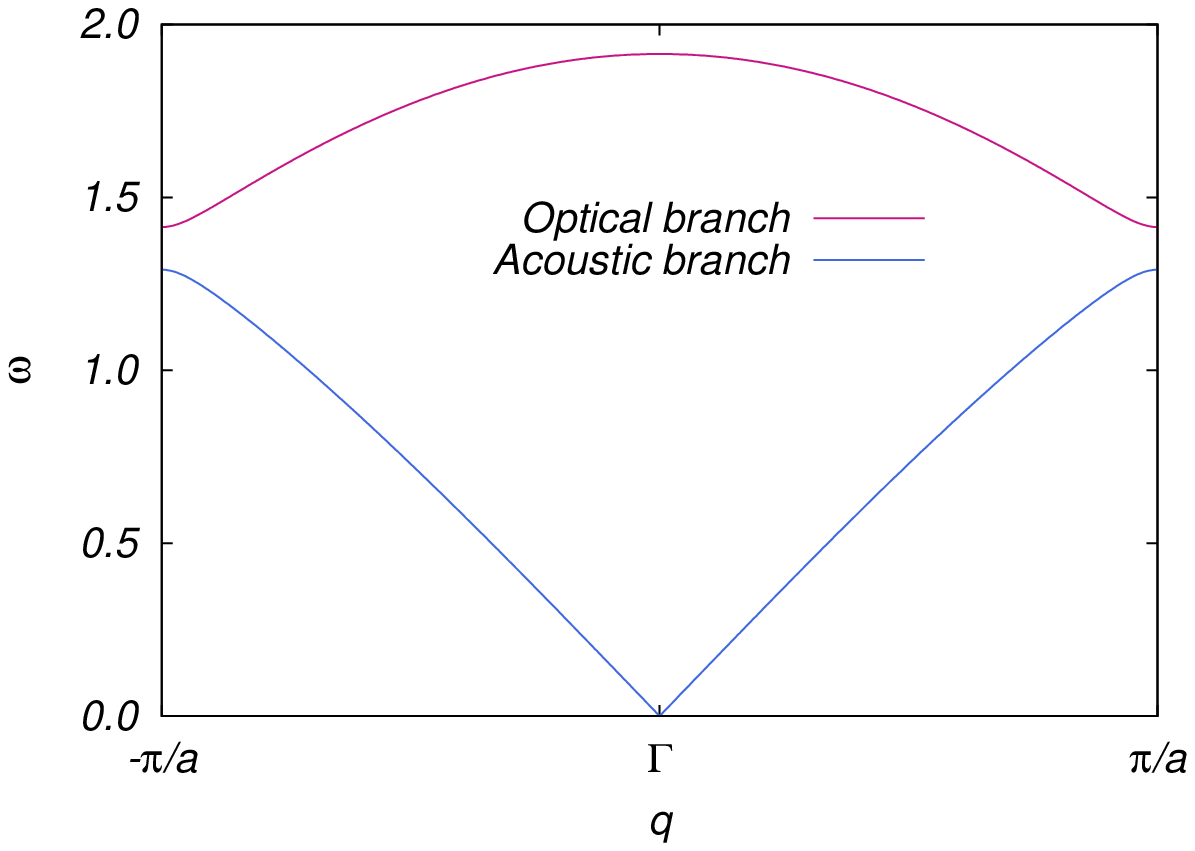}
		\caption{}
		\label{Phonondispersion}
	\end{subfigure}
	\begin{subfigure}[b]{.3\textwidth}
  		\centering
		\includegraphics[scale=0.37]{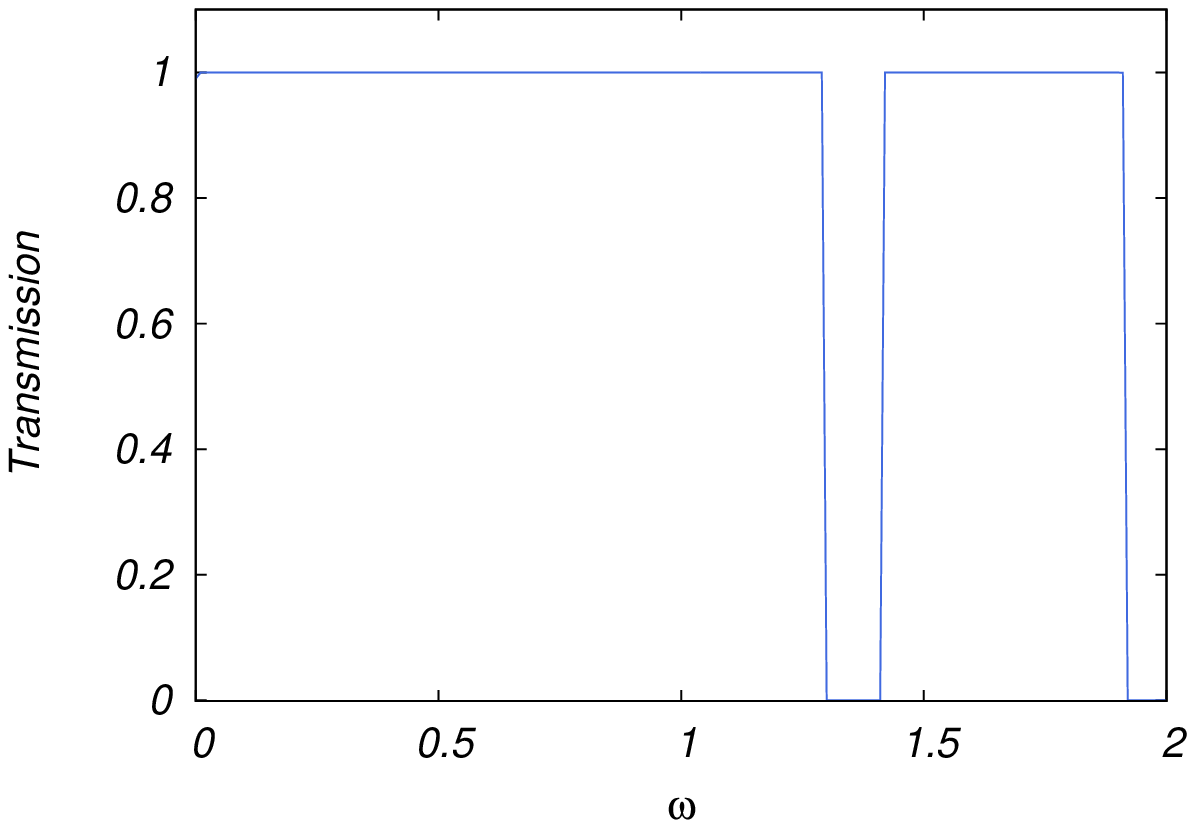}
		\caption{}
		\label{Phonontransmission}
	\end{subfigure}		
	\begin{subfigure}[b]{.3\textwidth}
  		\centering
		\includegraphics[scale=0.37]{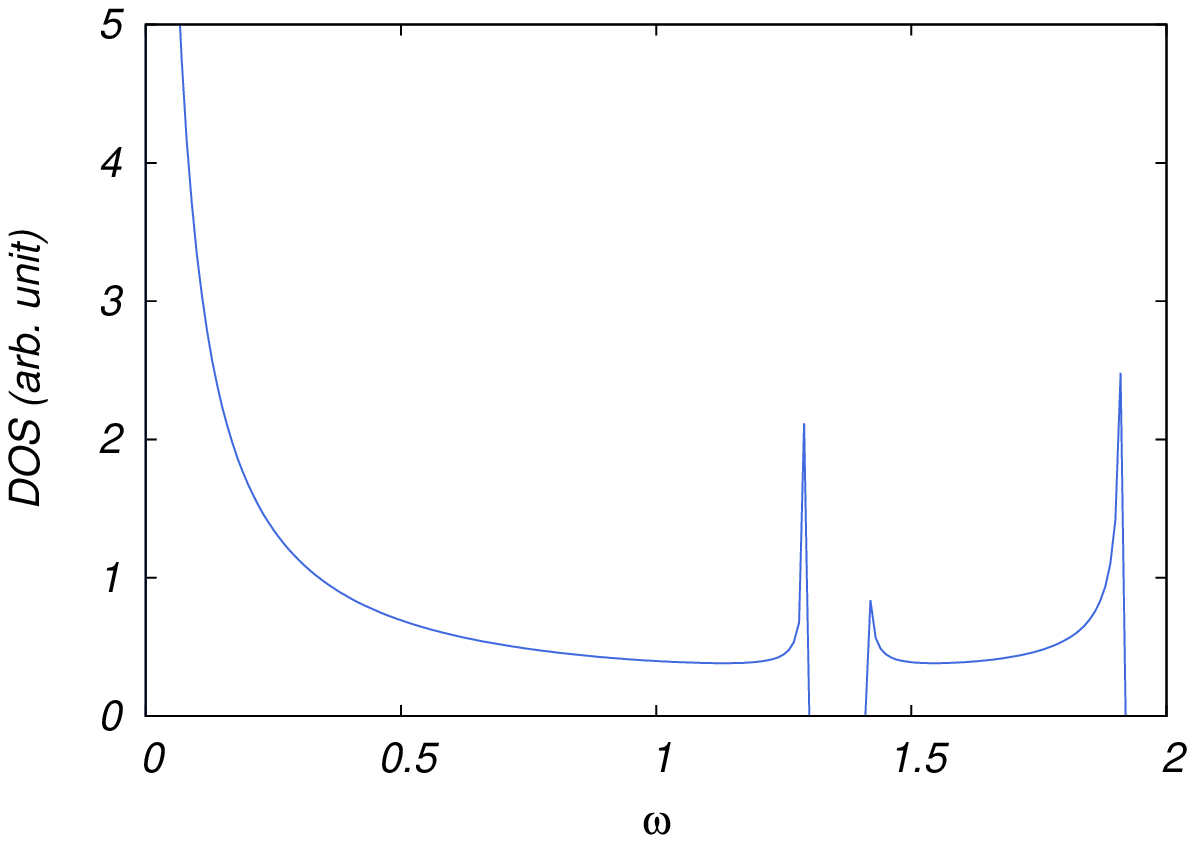}
		\caption{}
		\label{Phonondos}

	\end{subfigure}
\caption{(Colour online) Phononic properties of a typical chain consist of two different atoms as atomic basis: (a)Dispersion, (b)Transmission, (c)Density of states.}
\label{PhosphoreneProperties}
\end{figure}

To calculate phononic dispersion and transmission and density of states for a chain one needs to assemble a project just shown in figure \ref{NanoRibbonProjects}. The first calculating box contains dynamical matrix of the chain and the system attribute (Table \ref{TableTightbinding}) set on Nanoribbon for Dispersion box and set on Junction for Transmission box. As shown in figure \ref{PhononChainProjects}, Dispersion box connected to the Tight-binding box calculates dispersion and another box calculates phononic transmission and DOS. Figure \ref{Phonondispersion} shows acoustic and optical branches and figures \ref{Phonontransmission} and \ref{Phonondos} show the results of transmission and density of states for this project.

\section{Summary}
\label{Summary}
In summary, Condensed Matter Laboratory (CML) is introduced as a user-friendly application in the field of quantum computing simulation. The significant parts and important abilities of the CML were mentioned compactly and shown how much it is easy to use. In short, the most important methods we used in this application are the density functional theory and the Green's function theory in the tight-binding approximation. It is capable of simulating electronic and phononic properties of solids and nanostructures (ie. dispersion, transmission, density of states, current, etc). Also, it is able to calculate thermodynamic properties by means of statistical mechanics approaches. After that, some examples were presented to show the usages of the different parts of this application.



\end{document}